\documentstyle[epsf,12pt]{article}
\begin{document}
\begin{titlepage}
\title{Hierarchical population model\\ with a carrying capacity
distribution}
\author{J.O. Indekeu$^1$ and K. Sznajd-Weron$^2$
\\
$^1$Laboratory for Solid-State Physics and Magnetism\\ Katholieke
Universiteit Leuven, Leuven, Belgium\\ $^2$Institute for Theoretical
Physics\\ University of Wroclaw, Wroclaw, Poland\\ }
\maketitle
\begin{abstract}
A time- and space-discrete model for the growth of a rapidly
saturating local biological population $N(x,t)$ is derived from a
hierarchical random deposition process previously studied in
statistical physics. Two biologically relevant parameters, the
probabilities of birth, $B$, and of death, $D$, determine the
carrying capacity $K$. Due to the randomness the population depends
strongly on position, $x$, and there is a distribution of carrying
capacities, $\Pi (K)$. This distribution has self-similar character
owing to the imposed hierarchy. The most probable carrying capacity
and its probability are studied as a function of $B$ and $D$. The
effective growth rate decreases with time, roughly as in a Verhulst
process. The model is possibly applicable, for example, to bacteria
forming a ``towering pillar" biofilm. The bacteria divide on
randomly distributed nutrient-rich regions and are exposed to random
local bactericidal agent (antibiotic spray). A gradual overall
temperature change away from optimal growth conditions, for
instance, reduces bacterial reproduction, while biofilm development
degrades antimicrobial susceptibility, causing stagnation into a
stationary state.
\end{abstract}
\end{titlepage}

\setcounter{equation}{0}
\renewcommand{\theequation}{\thesection.\arabic{equation}}
\section{Introduction of the model}
As a concrete example, consider bacteria on a line with local
population $N(x,t)$ defined on $x \in [0,1]$. The population is
normalized, so that $N$ is to be interpreted as a density rather
than the number of bacteria. Generalization to two or three space
dimensions is straightforward, as will become clear from the simple
structure of the model. Assume, for simplicity, a homogeneous
initial condition $N_0 = N(x,0)>0$, representing inoculation along
the line of a thin uniform layer. At $t>0$ the growth process is
characterized by a probability of birth, $B$, which is applied to
segments or ``patches" along the line. The stochastic character of
the growth reflects that the surface is inhomogeneous. Some areas
are rich in nutrient (e.g., agar), others not. The division process
is effective only in nutrient-rich patches. These patches are
assumed to be of characteristic size $1/\lambda$, where $\lambda$ is
a scale factor central to the model. Alternative ways of inducing
spatially non-uniform growth or depletion are to apply, for
photosynthetic bacteria, favourable or unfavourable (UV)
illumination according to a structured or random spatial pattern
\cite{Neicu,Dahmen}.

Besides nutrient non-uniformity, the environmental conditions are
locally modified in a random way, for example by spraying drops of
an antibiotic which subsequently spreads by diffusion or transport
resulting from bacterial motility. In places with low chemical
concentration the population stagnates, elsewhere it shrinks. The
action of the bactericidal product is expressed through a
probability of death, $D$. An interesting variant on antibiotics,
and relevant in particular to inhibition of biofilm formation, is a
nutrient remover like, for example, lactoferrin. This protein, which
is abundant in human external secretions, mops up traces of iron and
thereby deprives bacteria of an essential substance. Applied in
concentrations below those that kill or prevent growth, it just
slightly increases the division time of the bacteria and enhances
their surface motility, causing them to ``twitch" or wander around
seeking food. As a result, microcolonies do not form and
matrix-encased communities specialized for surface persistence
(biofilms) do not develop \cite{Singh}.

The model is composed as follows. The characteristic size of the
spatial inhomogeneity of the nutrient is, for simplicity, taken to
be the same as that of the applied antimicrobial perturbation. The
(unit) line is thus divided in $\lambda$ segments, each of which has
the size of, say, an antibiotic drop. The probability of birth $B$
is applied to each segment, and subsequently the probability of
death, $D$. Each segment is thus visited twice. We have $0 \leq B,D
\leq 1$, and without loss of generality we take $\lambda = 3$ in
this paper.

One generation later this process is repeated, but the maximum local
increase or decrease of the population is reduced by a factor of
$\lambda$. This reduction is repeated in every generation. In this
way the growth process eventually comes to a halt, and the
population approaches a {\em carrying capacity}, as is commonly the
case in models for a single species \cite{Murray}. This rescaling of
the growth or the depletion is a crucial ingredient of the model and
leads to a hierarchical build-up of the population. While introduced
here heuristically, various justifications can be invoked for this
feature. It is known that bacterial reproductivity and/or
susceptibility to antimicrobial agents can ``wear out" in a
systematic manner, uniformly in space. For example, the {\em
effective birth rate} decreases significantly by varying the
temperature some 10 degrees away from optimal growth conditions.
Also, once a biofilm starts forming, it becomes resistant to
nutrient removers like lactoferrin, and it notoriously resists
killing by antibiotics.

Further, the rescaling by a factor $\lambda$ is applied not only to
the increase or decrease of the local population, but also to the
typical size of the nutrient-rich or bactericidal patches. The
justification for this runs as follows. Since after one generation
(of the order of two hours in real time) the nutrient or chemical
contamination has been redistributed in space, by diffusion or by
bacterial motion, regions which were initially rich (or pure) will
show poor (or toxic) spots, while initially barren regions will
display small fertile areas. Therefore, we apply the random birth
and death rules on a smaller length scale, reduced by a factor of
$\lambda$ with respect to the previous generation, and repeat this
reduction generation after generation.

In sum, every new generation the population fluctuations are reduced
by cooling (or heating) and biofilm development, and they occur on a
smaller length scale due to diffusive nutrient/antibiotic
fragmentation. A single scale factor controls - in this greatly
simplified model - all these dynamical complications, so that
neither external time-dependent fields nor ``interparticle"
interactions occur explicitly in what follows. Repeating this
process saturation is reached exponentially rapidly, typically in
less than 10 generations, as the calculations show. The resulting
local population density for long times, $N(x,\infty)$, is highly
inhomogeneous and largely reflects the initial random distribution
of the nutrient distribution and the toxic contamination, brought
about by spraying. Considered mathematically, $N(x,\infty)$ is a
fractal curve with infinite length, but trivial fractal dimension
$d_F=1$. Figure 1 shows an example of $N(x,5)$, the local population
after 5 generations, assuming $B=0.5$ and $D=0.2$.

\begin{figure}[htbp]
\centerline{\epsfxsize=8cm \epsfbox{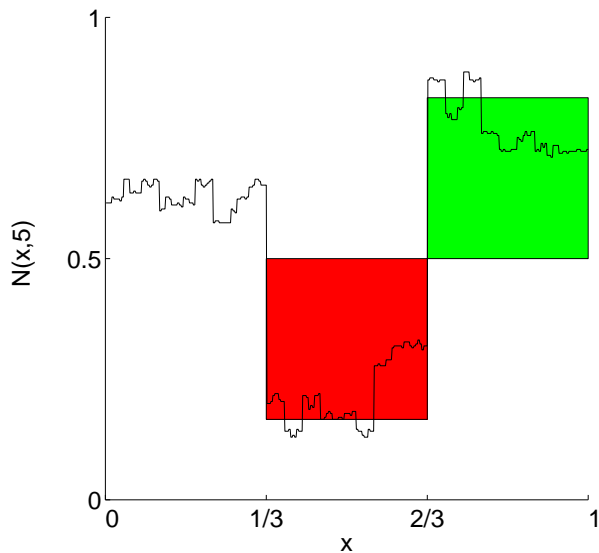}} \caption{Sample of a
local population density after 5 generations of hierarchical random
growth with $B=0.5$ and $D=0.2$, $N(x,5)$. The rescaling factor is
$\lambda =3$. The initial population is $N(x,0)=0.5$. The green
(red) area indicates the massive growth (depletion) which occurred
in the first generation. As a result, large jumps at $x=1/3$ and
$x=2/3$ persist for all later times.}
\end{figure}

Note that the natural death of the organisms is not included
explicitly in the model, for the simple reason that the induced
saturation occurs on a time scale (several hours) significantly
shorter than the natural lifetime. In other words, the probability
$D$ pertains to death caused by bactericidal products.

In Fig.1 it can be seen that what happens in the first generation
dominates the future evolution. For example, the green square
represents the massive growth on a nutrient-rich segment, $(2/3,1)$,
while the red square indicates the result of the killing of most of
the bacteria on a toxic patch, $(1/3,2/3)$. In the first segment,
$(0,1/3)$, nothing happened in the first generation. After a few
generations the population ``landscape" consists of towers and deep
crests, which get rougher as time progresses, but which hardly
change size anymore.

This model is mathematically equivalent to the previously introduced
hierarchical random deposition process \cite{IndFle} with rescaling
factor $\lambda$, and probabilities $P = B(1-D)$ for depositing a
hill, and $Q = D(1-B)$ for digging a hole, with $P+Q \leq 1$. The
correspondence between $(B,D)$ and $(P,Q)$ can be seen by taking
into account that in the biological application every segment is
visited twice, applying $B$ in the first visit followed by $D$ in
the second, whereas in the deposition model there is only one visit
per segment, per generation. Therefore, after a complete visit the
population is increased provided offspring was produced {\em and} no
deaths occurred, whence the product $B(1-D)$. The population is
decreased provided no divisions occurred {\em and} the antibiotics
killed a fraction of the existing population, whence $(1-B)D$.
Finally, the population remains constant if {\em either} offspring
was produced but subsequently killed, {\em or} no births took place
and the antibiotics were absent or had no effect. The probability
for this is $BD+(1-B)(1-D)= 1-(P+Q)$. Note that the total
probability for the three outcomes is unity.

Using the deposition model it is easy to derive analytical results
in terms of the variables $P$ and $Q$, and also numerical simulation
is straightforward. In generation $n$ the process requires just
$\lambda^n$ (pseudo-)random numbers. For each segment of width
$\lambda^{-n}$ along $x$ the population is increased by an amount
equal to the segment size (in dimensionless units) with probability
$P$, or reduced by that amount with probability $Q$, or left
unchanged with probability $1-(P+Q)$. This leads to the
characteristic rapidly converging population landscape formed by
squares of decreasing size (Fig.1).

The initial condition is taken to be $N_0 = 1/(\lambda -1)$,
independent of $x$ (uniform inoculation). The first quantity of
interest is the mean carrying capacity, $\bar K$, which is the
ensemble average of $N(x,\infty)$ over all possible random
realizations of the process. Since on average the population grows
by an amount $(B-D)\lambda^{-n}$ in generation $n$, we obtain
\begin{equation}
\bar K = (1+ B-D)/(\lambda -1) = (1+P-Q)/(\lambda -1),
\end{equation}
independent of $x$. Clearly, the initial condition $N_0$ has been
chosen large enough so that, for all $B$ and $D$, the population
remains positive for all $x$ and $t$, and approaches zero only in
the extinction limit $B=0$, $D=1$. Note the simple identity
$B-D=P-Q$, making the relations between the growth probabilities and
the deposition parameters more transparent.

Generalization of the model to higher dimensions (e.g., substrate
dimension 2) is not pursued here, but the results of such extension
can easily be anticipated by examining the landscapes shown for the
hierarchical deposition model \cite{IndFle,IFPB} on planar
substrates. In fact, if the bacterial population is compact, the
value of $N(x,t)$ is simply proportional to the height of the
biofilm above the substrate, and the landscapes can be interpreted
as three-dimensional in real space. This correspondence holds as
long as there are no ``overhangs" in the biofilm, and thus works for
towering pillar structures but not for mushroom-shaped ones
\cite{Singh}. The similarity between the three-dimensional
characteristic landscapes generated in this model, and the structure
of its horizontal cross-section\cite{IFPB} (see Fig.6 in that
reference), and images of vertical and horizontal sections of a
towering pillar biofilm, as obtained by confocal laser scanning
microscopy, is remarkable\cite{Herma}. This suggests that a
hierarchical model may be a reasonable first approximation for
describing biofilm growth.

The connection between the model and the familiar Verhulst process
is elucidated in Section 2. It is made clear that $B$ and $D$ are
biologically relevant parameters determining the carrying capacity
$K$, whereas $\lambda$ simply sets an overall time scale factor. The
carrying capacity distribution, the key result of this random growth
model, is calculated analytically in Section 3, and concrete
examples are discussed in Section 4, together with a check against
numerical simulation. The self-similar character of this
$K$-distribution is demonstrated and the effect of $B$ and $D$ on
the standard deviation of the population is examined in
representative cases. Section 5 deals with the most probable
carrying capacity and its probability, considered as a function of
$B$ and $D$. Insight is gained in the abundance of large deviations
away from the most probable $K$. The final section, 6, presents our
conclusions and addresses a possible experimental test of this
model.

\setcounter{equation}{0}
\renewcommand{\theequation}{\thesection.\arabic{equation}}
\section{Connection with a Verhulst process or logistic growth}
In this section we are concerned with the uniform, i.e., spatially
averaged, population. The time evolution of the mean, or ``quenched
average", of $N(x,t)$ is given by the change in the mean value,
$\bar N(t)$, from $t=n$ to $t=n+1$,
\begin{equation}
\bar N(n+1) = \bar N(n) + \lambda^{-n-1}(B-D),
\end{equation}
which is solved by
\begin{equation}
\bar N(n) = N_0 + (1-\lambda^{-n})(B-D)/(\lambda -1)
\end{equation}
This is a simple growth, converging exponentially rapidly, for $n
\rightarrow \infty$, to the mean carrying capacity. It is useful now
to work with the differences
\begin{equation}
\Delta \bar N = \bar N -N_0, \,\;\;\mbox{and}\;\;\, \Delta \bar K =
\bar K -N_0,
\end{equation}
with respect to the initial population $N_0 = 1/(\lambda - 1)$. We
obtain
\begin{equation}
\Delta \bar N(n) = (1-\lambda^{-n})\Delta \bar K
\end{equation}
Using this in (II.1) leads to
\begin{equation}
\Delta \bar N(n+1) - \Delta \bar N(n) = \frac{\lambda -1}{\lambda}
\frac{\lambda^{-n}}{1-\lambda^{-n}} \Delta \bar N(n)
\end{equation}
For large $n$, we can neglect $\lambda^{-n}$ compared to 1 in the
denominator, and again using (II.4), we get the {\em nonlinear}
difference equation
\begin{equation}
\Delta \bar N(n+1) - \Delta \bar N(n) \approx \frac{\lambda
-1}{\lambda}\Delta \bar N(n) (1- \frac{\Delta \bar N(n)}{\Delta \bar
K})
\end{equation}
This is clearly a discrete Verhulst process, the continuum time
limit of which is of the form of the Verhulst equation
\begin{equation}
\frac{d\Delta \bar N(t)}{dt} = \frac{\lambda -1}{\lambda}\Delta \bar
N(t) (1- \frac{\Delta \bar N(t)}{\Delta \bar K})
\end{equation}

This analogy allows us to identify and interpret the model
parameters unambiguously. The characteristic time scale of the model
is $\lambda/(\lambda-1)$, a trivial constant. Therefore, the length
rescaling factor $\lambda$ in the model has no particular biological
relevance, and can be given an arbitrary integer value $\lambda >
1$. The carrying capacity (difference) given by $\Delta \bar K =
(B-D)/(\lambda -1)$ is an important property, and we find that it is
determined simply by the difference of the birth and death rates.
This link between a phenomenological parameter of the Verhulst
equation, the carrying capacity, and the stochastic ``microscopic"
model parameters $B$ and $D$ is a key ingredient of our model.
Finally, the effective birth rate $1/\tau$ depends on the size of
the population, and eventually vanishes as the carrying capacity is
approached, as is typical of logistic growth. Asymptotically for
long times, it is given by
\begin{equation}
1/\tau \sim \frac{\lambda -1}{\lambda}(1- \frac{\Delta \bar
N(t)}{\Delta \bar K}) =\frac{\lambda -1}{\lambda}\lambda^{-n}
\end{equation}
More precisely, the exact expression for all times is
\begin{equation}
1/\tau =\frac{\lambda
-1}{\lambda}\frac{\lambda^{-n}}{1-\lambda^{-n}}
\end{equation}
Note, and we wish to stress, that for our process the Verhulst
equation is nothing but an easily interpretable {\em nonlinear}
approximation to the actual linear differential equation describing
our dynamics,
\begin{equation}
\frac{d\Delta \bar N(t)}{dt} = \frac{\lambda -1}{\lambda}\Delta \bar
N(t) (1- \frac{\Delta \bar N(t)}{\Delta \bar K})\frac{\Delta \bar
K}{\Delta \bar N(t)} = \frac{\lambda -1}{\lambda} (\Delta \bar
K-\Delta \bar N(t))
\end{equation}

The precise way (in our case exponential) in which the effective
birth rate vanishes is only one possible way in which this rate in a
biological system can be driven away from its optimal value. We
envisage that this drift is achieved by a temperature variation,
upwards or downwards, bringing the bacterial growth to a halt on a
time scale of a small number of generations. In the introduction we
already alluded to the fact that biofilm development can be a factor
leading to resistance to antibiotics, which can lead to a vanishing
death rate. We simply assume that both effects (temperature change
and biofilm formation) are present uniformly throughout the sample.
Our justification for assuming a uniform external perturbation
responsible for the saturation is that in our random model the
effective birth rate (II.9) does not depend on the local
($x$-dependent) population, but only on time. Alternatively, the
self-limitation of the population could of course also be due to
{\em local} overpopulation, but this model is in its present form
too simple to allow for an internal local feed-back process.

\setcounter{equation}{0}
\renewcommand{\theequation}{\thesection.\arabic{equation}}
\section{The carrying capacity distribution}
Even though the average growth of the population is characterized by
trivial exponential saturation, the local population $N(x,t)$ for a
particular realization of the quenched randomness typically shows
interesting large fluctuations. These deviations, highly nonuniform
in space, can possibly be measured experimentally by probing the
local population of bacteria after spraying a few drops of an
antibiotic. The fluctuations of the total population ${\cal N}(t)=
\int_0^1 N(x,t) dx$ are studied in this section and the next two.

Our aim in this section is to explore the effect of random nutrient
inhomogeneity and random antibiotic spray on the carrying capacity
$K$, which is the asymptotic value for long times of the total
population. The simple stochastic nature of the model allows to
obtain the full $K$-distribution in analytic form in terms of the
probability of birth $B$ and that of death $D$, with modest
mathematical effort. Consider a particular random evolution. In
generation $m$ the total population grows by an amount
$(\lambda^m-k_m)/\lambda^{2m}$ where the possible values of the
integer $k_m$ are contained in the set $ \{ 0,1,...,2\lambda^{m}
\}$. For example, in the deterministic limits $B=1$ and $D=0$, one
always has $k_m = 0$, and for $B=0$ and $D=1$ one invariably
encounters $k_m = 2\lambda^{m}$. The set $\{k_1, ..., k_n \}$ will
be referred to as a ``growth sequence" of length $n$. Of course,
growth sequences are in general degenerate in the sense that
different population landscapes can possess the same growth
sequence. The total population after $n$ generations, denoted by
${\cal N}(n)$ is then given by
\begin{equation}
{\cal N}(n) = \frac{1}{\lambda -1} + \sum _{m=1}^{n} \frac{\lambda^m
-k_m}{\lambda^{2m}}
\end{equation}
At this point it is useful to note that different growth sequences
may accidentally result in the same value for the total population.
For example, a single local increase of $N$ in one generation can be
deleted by $\lambda^2$ local decreases in the next generation.
Therefore, an alternative sequence in which nothing changes in those
two generations would lead to the same total population.

If we work with the deposition probabilities, related to $B$ and $D$
by $P=B(1-D)$ and $Q= D(1-B)$, and define $S = P+Q$, the probability
for depositing in generation $m$ any configuration with, say, $h$
holes, $f$ flat segments (neither hill nor hole), and $\lambda^m
-f-h$ hills is given by the following expression which takes into
account the number of ways in which the holes and flat segments can
be put in the standard combinatorial way,
\begin{equation}
{\cal P}_m(h,f) =  \left ( \begin{array}{c} \lambda^m \\ h
\end{array} \right )
\left ( \begin{array}{c} \lambda^m -h \\ f
\end{array} \right ) P^{\lambda^m - f - h} (1- S)^{f}
Q^h
\end{equation}
Note that summing this expression over all possible $h$ and $f$
gives the total probability,
\begin{equation}
\sum_{h=0}^{\lambda^m} \sum_{f=0}^{\lambda^m-h} {\cal P}_m(h,f) =  1
\end{equation}
which is unity, as it should.

The probability for having a fixed population increment, that is, a
particular value $k_m=k$, in generation $m$, is given by the
following partial sum over the ${\cal P}_m(h,f)$,
\begin{equation}
\hat{\cal P}_m(k) = \sum _{h=j(k)}^{[k/2]} \left ( \begin{array}{c}
\lambda^m \\ h
\end{array} \right )\left ( \begin{array}{c} \lambda^m -h \\ k-2h
\end{array} \right )  P^{\lambda^m - k + h} (1- S)^{k-2h} Q^h
\end{equation}
Note that $k$ equals the number of flat segments plus twice the
number of holes, so that $f=k-2h$. Here, $[k/2]$ is equal to $k/2$
if $k$ is even, and equal to $(k-1)/2$ if $k$ is odd. Further, the
integer $j(k)$ equals 0 if $k \leq \lambda^m$, and $j(k) = k-
\lambda^m$ for $k > \lambda^m$. One can verify the normalization
\begin{equation}
\sum_{k=0}^{2\lambda^m} \hat{\cal P}_m(k) = 1
\end{equation}

Since the $k_m$, for different generations $m$, are independent
variables, the probability of a given growth sequence is the product
\begin{equation}
\hat \pi_n({\bf k}) \equiv \prod _{m=1}^{n} \hat{\cal P}_m(k_m),
\end{equation}
where ${\bf k} = (k_1, k_2, ..., k_n)$. Consequently, the
probability $\Pi$ of realizing the total population ${\cal N}(n)$ is
the sum of the probabilities of all the growth sequences ${\bf k}$
which produce the same value ${\cal N}(n)= {\cal N}$,
\begin{equation}
\Pi_n({\cal N}) = \sum _{\bf k}^{({\cal N})} \hat \pi_n({\bf k}),
\end{equation}
where the sum takes care of the degeneracy of ${\cal N}$. In
practice this sum contains only a few terms, because typically not
many growth sequences lead to the same value for the total
population. For large $n$ the populations ${\cal N}(n)$ converge to
the carrying capacities $K$, so that the distributions $\Pi_n ({\cal
N})$ converge to the {\em carrying capacity distribution}
$\Pi_{\infty}(K)$, or in short $ \Pi (K)$, which we set out to
obtain.

Note that the total probability for the entire process up till any
given generation $n$ is correctly normalized, since
\begin{equation}
1 = \prod _{m=1}^{n} \sum_{k=0}^{2\lambda^m} {\hat{ \cal P}}_m(k) =
\sum_{\bf k} \prod_ {m=1}^{n} {\hat{ \cal P}}_m(k_m) = \sum_{\bf k}
\hat \pi _n ({\bf k})= \sum_{\cal N} \Pi _n ({\cal N})
\end{equation}
In particular,
\begin{equation}
\sum_K \Pi(K) = 1
\end{equation}

In closing this section a few technical remarks are in order.
Clearly, in the special cases $P=0$, $Q=0$, or $P+Q=1$ the trinomial
formulae we have derived, in particular (III.2) and (III.4), cannot
be used in the present form. Instead, simple binomial expressions
should be written down in those cases. Further, in view of the
relations $P=B(1-D)$ and $Q=D(1-B)$ one can verify that the
``biological" domain in the $(P,Q)$-plane occupies only the
kite-shaped region below the line $\sqrt{P}+\sqrt{Q}=1$. On this
line we have $B=1-D$. We will discuss later that this is a symmetry
line in the $(B,D)$-plane. Therefore, the present biological
application uses only a subset of the parameters of the physical
deposition model. This is illustrated in Figure 2. The relevant
kite-shaped region is filled in black. The white area below $P+Q=1$
is not accessible in the population model, but it is available in
the deposition model.

\begin{figure}[htbp]
\centerline{\epsfxsize=10cm \epsfbox{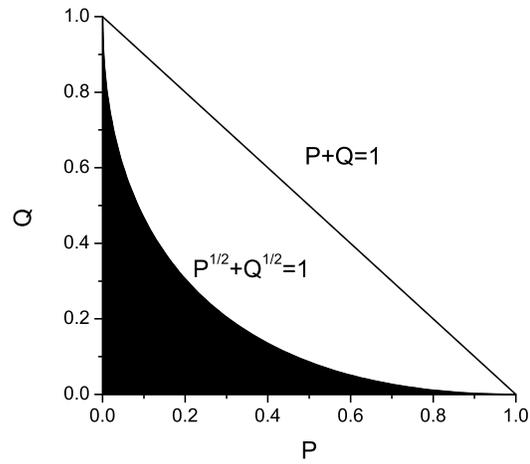}} \caption{The dark
area in the $(P,Q)$-plane below the line $\sqrt{P}+\sqrt{Q}=1$
corresponds to the ``biologically" accessible range of $P$ and $Q$,
which can be reached starting from probabilities $0 \leq B, D \leq
1$ through the relations $P=B(1-D)$ and  $Q = D(1-B)$. Note that
$P+Q \leq 1$ gives an upper bound on the available ``physical" range
of $P$ and $Q$ in the deposition model.}
\end{figure}

\setcounter{equation}{0}
\renewcommand{\theequation}{\thesection.\arabic{equation}}
\section{Bacterial population distributions versus antibiotic efficiency}
In order to illustrate the carrying capacity distribution and its
properties we turn to a few concrete examples. Consider a process
with $B=0.5$, representing equally probable nutrient-rich and
nutrient-poor patches, and $D=0.5$, corresponding to medium
antibiotic strength. It follows that $P= 0.25$ and $Q= 0.25$. Figure
3a shows the population distribution after one generation,
$\Pi_1({\cal N})$, for $\lambda =3$. Since in the first generation
the population is uniquely determined by the growth sequence $k_1$,
through ${\cal N}(1) = N_0+ (\lambda -k_1)/\lambda^2$, with $k_1= 0,
..., 2\lambda$, we have
\begin{equation}
\Pi_1({\cal N}) = \pi_1(k({\cal N}))
\end{equation}
The probabilities associated with the seven peaks in the figure thus
follow directly from (III.4). Note that the distribution is
symmetric about the initial population $N_0=0.5$, due to the fact
that $B=D$.

\begin{figure}[htbp]
\centerline{(a) \epsfxsize=8cm \epsfbox{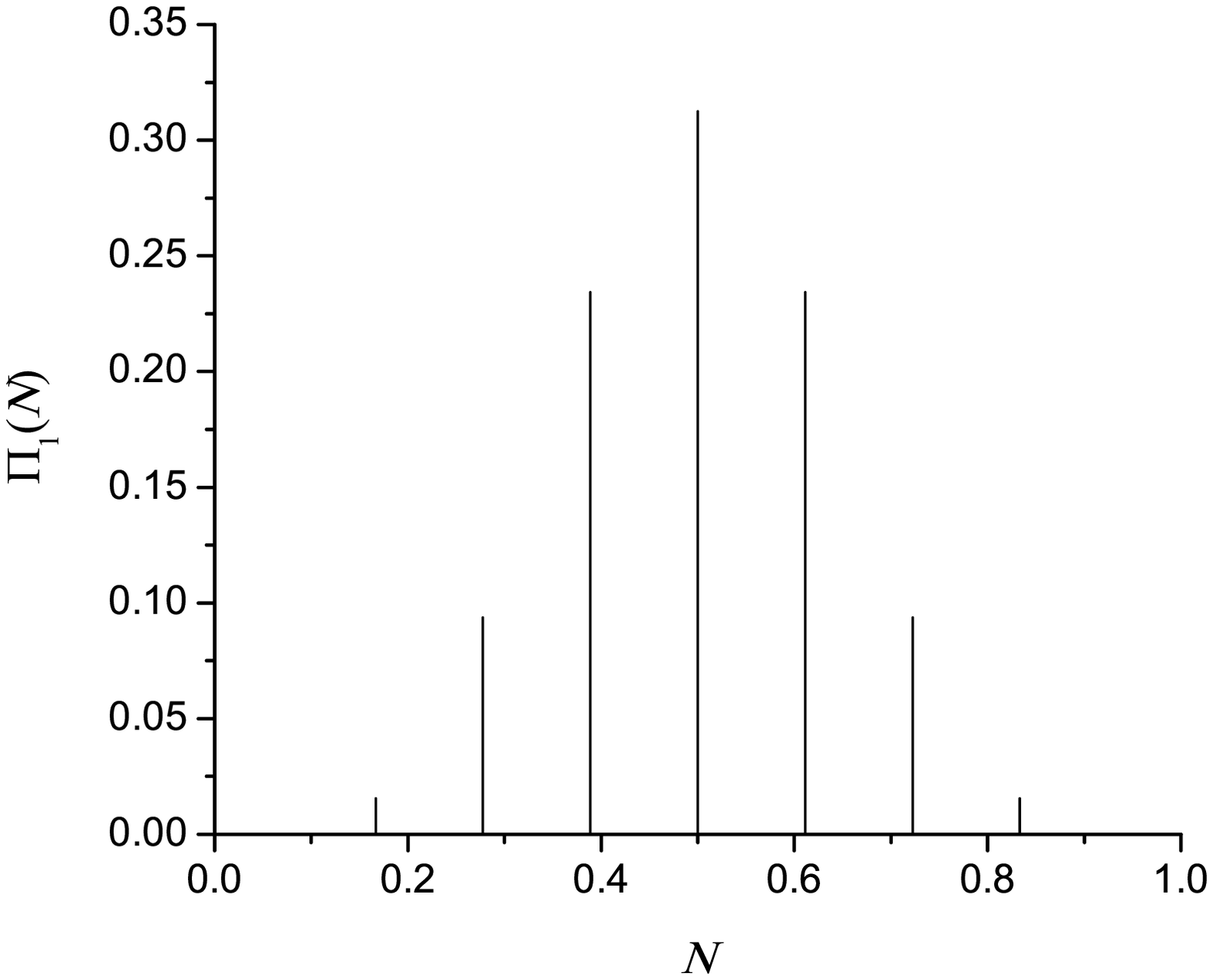}
            (b) \epsfxsize=8cm \epsfbox{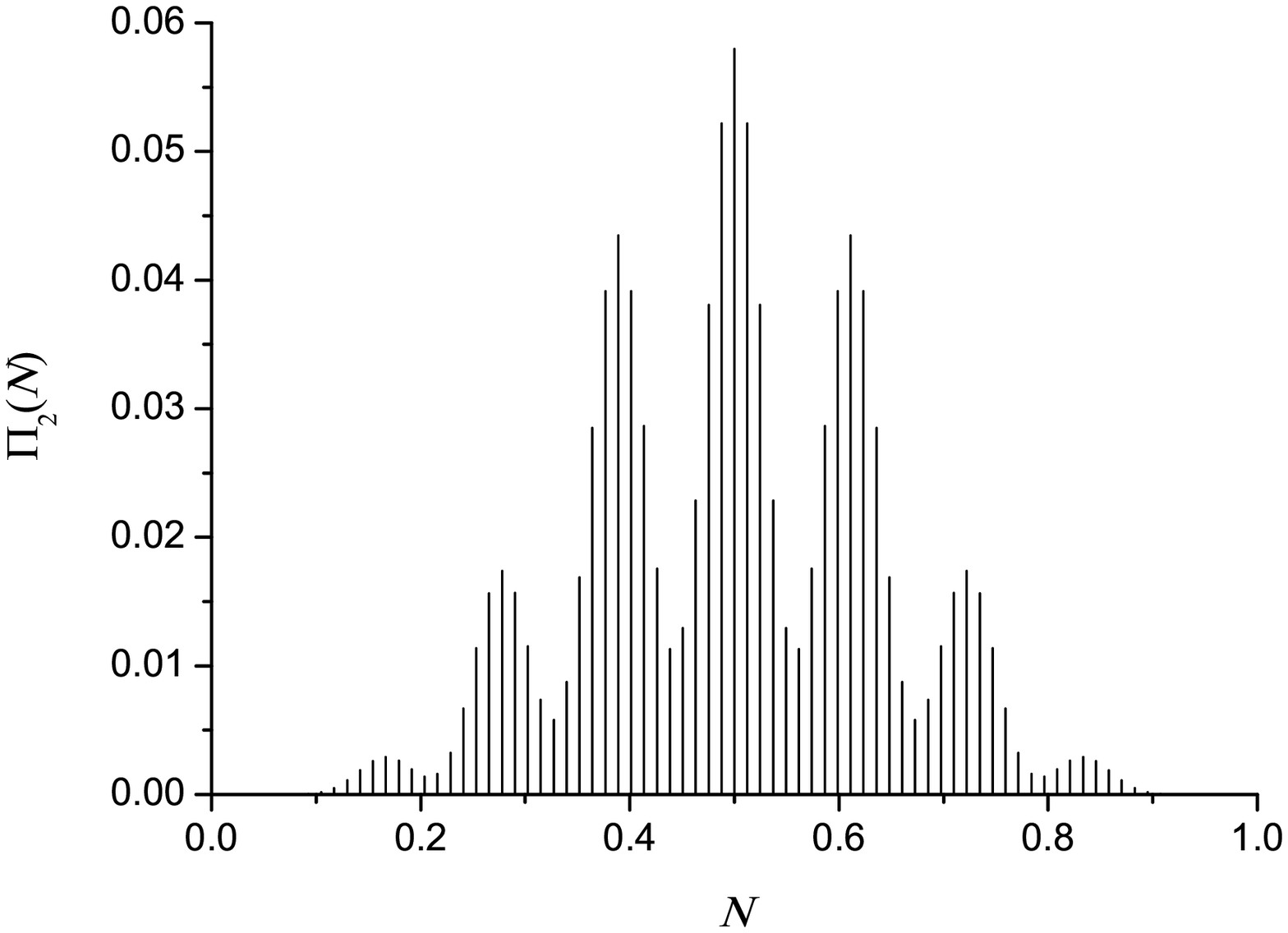}}
\centerline{(c) \epsfxsize=10cm \epsfbox{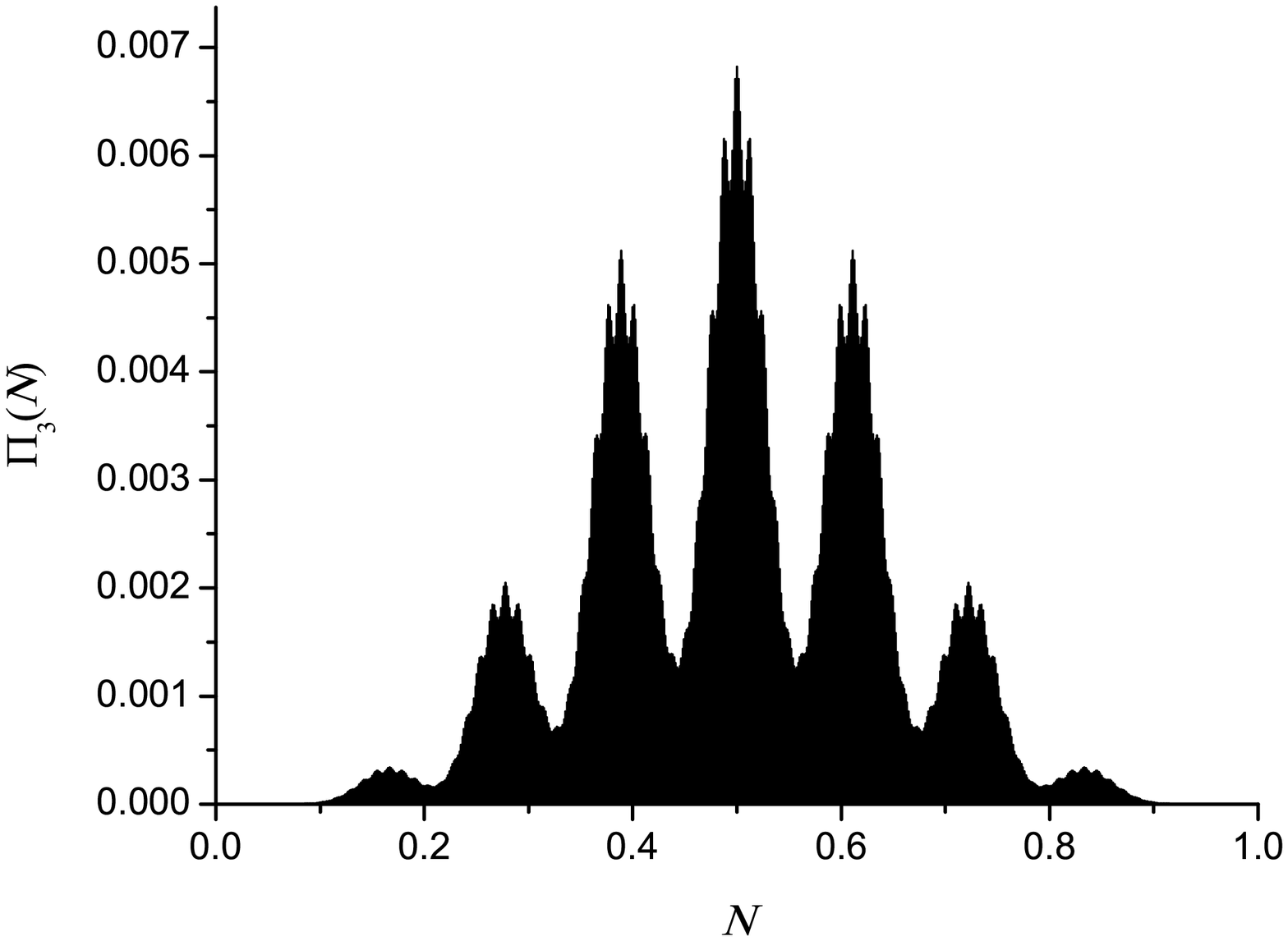}}
\caption{
(a)Population distribution after one generation,
$\Pi_1({\cal N})$, for $B=0.5$ and $D=0.5$. The most probable
population (highest peak) coincides with the initial population.
(b)Population distribution after two generations,
$\Pi_2({\cal N})$, for $B=0.5$ and $D=0.5$. This distribution arises
as the convolution of two generators. The seven peaks originate from
the seven populations seen in Fig.3a.
(c)Population distribution after three generations,
$\Pi_3({\cal N})$, for $B=0.5$ and $D=0.5$. Notice the
self-similarity in the fine-structure of the main peaks. The
distribution, normalized to unity, gives the probabilities of 703
distinct population values and its shape (but not its height) has
almost converged to that of the long-time carrying capacity
distribution $\Pi(K) \equiv \Pi_{\infty}({\cal N})$.
}
\end{figure}

In the language of fractal geometry \cite{Mandelbrot} Fig.3a can be
called the ``generator" of the distribution. If this generator were
now applied on a smaller scale to split every peak into 7 new peaks,
we would obtain a perfect self-similar object, when iterated ad
infinitum. However, in the second generation the generator is
different and has $2\lambda^2 + 1$ peaks, 19 in our example. Its
width is reduced by a factor of $\lambda$, as can be seen from
inspection of (III.1). It is the convolution of the two generators,
according to the product expression (III.6) and the sum (III.7),
which gives the population distribution after two generations. The
result is shown in Fig.3b.

In this distribution the 7 probability peaks of the first generation
are still clearly visible, but they have been split due to the
relatively small population shifts obtained in the second
generation. The 7 structured peaks thus have a width and fine
structure determined by the generator of the second generation, and
all 7 are similar to each other, differing only in their amplitude
which is apparent in Fig.3a. Note that the total number of peaks,
73, is significantly smaller than the product $7 \times 19$. This is
due to overlap of peaks, which results, as discussed previously,
from different growth sequences leading to the same population.

As the generations progress, the individual probability peaks can no
longer be resolved on the normal population scale. This is the case
already after three generations, as Fig.3c shows. There are 703
peaks after three, and 6481 peaks after four generations. The global
scaling properties of the distribution can be easily understood if
one takes into account that the number of peaks increases by
approximately a factor of 9, since the inter-peak distance, or
minimum population shift, decreases by a factor of $\lambda^2$ in
each generation, as implied by (III.1). Consequently, in order to
preserve the normalization of the total probability the height of
the peaks must decrease roughly by a factor of 9 also.

After four generations the population distribution looks nearly
identical to that after three generations, provided the height is
scaled by roughly a factor of 9. Therefore, Fig.3c (three
generations) already reveals the shape of the population
distribution in the limit of a large number of generations, i.e., it
gives the carrying capacity distribution $\Pi (K)$ that we are
interested in.

Besides the analytic calculation, we have carried out simulations of
these random processes and obtained population distributions in the
form of histograms. Obviously, for a histogram to present accurately
a discrete distribution of peaks, the bin width must be smaller than
the inter-peak separation in population. For example, from these
simulations, typically involving $10^4$ processes for given $B$ and
$D$, a histogram with 1000 bins on $(0,1)$ is obtained which
reproduces the analytic results accurately. In view of the fact that
little or no changes can be perceived in the shape of the (properly
scaled) distribution for generation numbers larger than three, it
follows that a histogram with fixed bin size (1000) will converge
(without scaling) already after three or four generations and reveal
the carrying capacity distribution. In actual experiments a bin size
is defined by practical resolution limitations on the counting of
bacteria. Therefore, the biologically relevant frame-work is that
which employs a {\em fixed bin size}.

The example we have chosen in Figs.3a-c is a typical case in the
category of broad carrying capacity distributions with an average
value close to (in this case at) the initial homogeneous population
$N_0$. For medium nutrient levels and medium antibiotic strength the
fluctuations are apparently too large to gain ``control" over the
outcome of the experiment. The situation changes drastically when
bacteria in a nutrient-poor environment are exposed to strong
antibiotics. In this case a thorough elimination of bacteria results
almost certainly. This is shown in Fig.4.

\begin{figure}[htbp]
\centerline{\epsfxsize=10cm \epsfbox{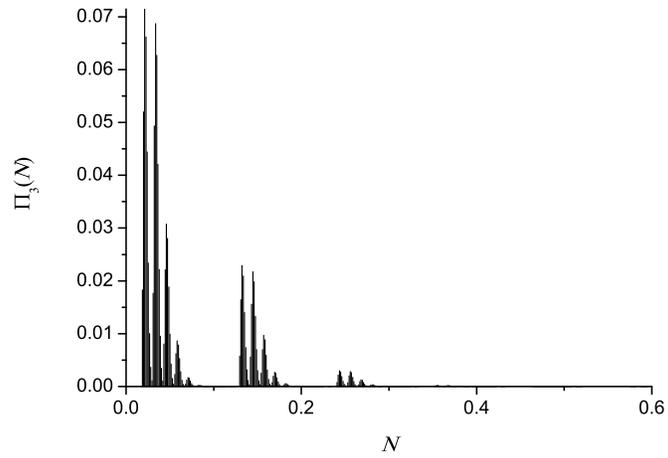}}
\caption{Population distribution after three generations,
$\Pi_3({\cal N})$, for $B=0.05$ (poor nutrient) and $D=0.95$ (strong
antibiotics). The population, initially at ${\cal N} = 0.5$ is
strongly suppressed to low values and with small standard deviation.
The mean value, 0.067, calculated from (II.2), is slightly larger
than the most probable value.}
\end{figure}

For birth and death probabilities $B=0.05$ and $D=0.95$, Fig.4 gives
the carrying capacity distribution after three generations. This
corresponds to $P=0.0025$ and $Q= 0.9025$. Clearly, the population
has been suppressed to far below its initial value $N_0=0.5$, and
the probability is concentrated on the lowest accessible population
values, around ${\cal N} \approx 0.06$. This result is to be
expected, and in line with common experience with the effect of
antibiotics on bacteria in a nutrient-poor medium.

Possibly a more surprising result from this model is what happens
when medium nutrient level is combined with strong (or very strong)
antibiotics. If we take $B=0.5$ and $D=0.99$, leading to $P=0.005$
and $Q= 0.495$, we obtain the carrying capacity distribution shown
in Fig.5. The following properties are conspicuous: i) the
population is entirely shifted to well below its initial value
$N_0=0.5$, ii) the distribution is broad; iii) the average
population is about half the initial population. It appears that the
antibiotics has failed in two respects. The population has not been
suppressed completely, and it is difficult to predict the eventual
asymptotic population due to the large fluctuations. Mathematically,
this insufficiency can be traced to be caused, not by the medium
abundance of nutrient-rich patches in itself, since only very little
growth is observed ($P=0.005$ is very small), but by a substantial
reduction of the antibiotic efficiency. This reduction, from
$D=0.99$ in biological parameter space, to $Q=0.495$ in physical
parameter space, is due to the {\em indirect} effect of
nutrient-rich spots on antibiotic strength through $Q=D(1-B)$.
Indeed, on $50 \%$ of the biofilm the antibiotic efficiency is spent
on taking out only the newly produced bacteria, since $B=0.5$, and
only on the remaining $50\%$ of the area the initially existing
population gets (almost) killed eventually, so that we can
understand that $\bar N (n) \rightarrow 0.255$, for $n \rightarrow
\infty$.

\begin{figure}[htbp]
\centerline{\epsfxsize=10cm \epsfbox{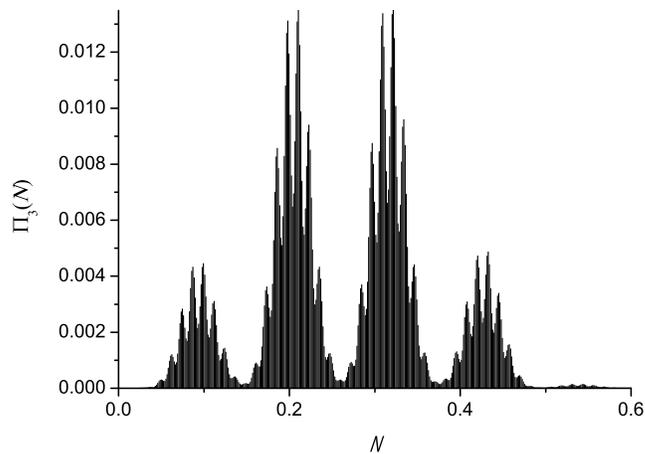}}
\caption{Population distribution after three generations,
$\Pi_3({\cal N})$, for $B=0.5$ and $D=0.99$. In spite of the strong
antibiotics the nutrient-rich surface allows the population to
survive at, on average, half the initial value, and with large
fluctuations. Notice that the mean value, 0.264, calculated from
(II.2), is roughly in between the two most probable values.}
\end{figure}

The few representative cases we have discussed up to now are the
most interesting ones which can be studied within this very simple
model. We do not need to devote special attention to the limits $B
\approx D \approx 0$ or, equivalently, $B\approx D \approx 1$, for
which the distribution remains sharply concentrated around the
initial value, so that the outcome of experiments is simple to
predict. In the next section we elaborate on the behaviour of the
most probable carrying capacity and its probability as a function of
the birth and death rates, i.e., of $B$ and $D$.

\setcounter{equation}{0}
\renewcommand{\theequation}{\thesection.\arabic{equation}}
\section{Most probable carrying capacity and its probability}
With regard to an actual experiment it is important to predict the
{\em most probable} carrying capacity, $K^*$, that will be observed,
and its probability of occurrence, $\Pi (K^*)$, given the birth and
death probabilities $B$ and $D$. In general $K^*$ differs from the
mean $\bar K$. We have studied these properties of the model for
rescaling factors $\lambda =2$ and $\lambda = 3$. Differences in
rescaling factor are seen to lead to quantitative differences only,
not essential for our discussion. Here we report on the results for
$\lambda=3$.

Since the population distributions converge rapidly with increasing
generation number, the results for the first generation (the
``generator") determine to a very good approximation the relative
importance of the peaks in the final distributions, as well as the
population values associated with the peaks. Therefore, we can
restrict our attention to the {\em first generation} when studying
the most probable carrying capacity and its probability {\em
qualitatively}, and work with the most probable population ${\cal
N}^*$ and its probability $\Pi_1({\cal N}*^)$. In Appendix A we give
the analytic expressions for the probabilities of the populations
after one generation.

There are two lines of symmetry in the model, which merit separate
attention. The line $B=D$ is a symmetry line in the sense that the
population distributions for $B>D$ are the mirror images, reflected
about $N=0.5$, of the distributions with $B$ and $D$ interchanged.
On the line $B=D$ the average population remains equal to the
initial population. There is no net growth (see Section II). One
would therefore suspect that, when the antibiotic strength precisely
compensates the nutrient abundance, the most probable population
after the experiment equals the initial population. This is indeed
the case. However, unless $B$ is sufficiently close to zero or to 1,
the fluctuations are large and the probability $\Pi_1(0.5)$ of
observing the most probable population is less than $50 \%$. This is
illustrated in Fig.6 (case $B=D$).

\begin{figure}[htbp]
\centerline{\epsfxsize=8cm \epsfbox{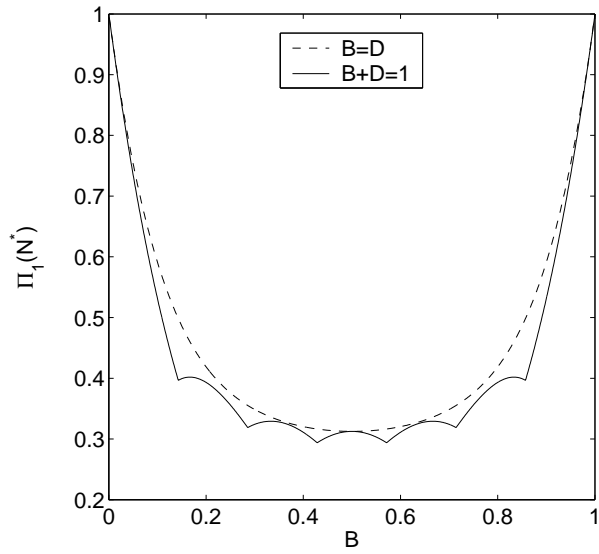}}
\caption{Probability of the most probable population after
one generation, as a function of $B$, for the special symmetry lines
$B=D$ (dashed line without singularities; along this line ${\cal
N}^* = \bar N = N_0 =  0.5$) and $B+D=1$ (solid line with six corner
singularities).}
\end{figure}

The second symmetry concerns the substitution of $B$ by $1-D$, and
of $D$ by $1-B$. Clearly, this operation leaves $P$ and $Q$
unchanged, and therefore the results are invariant. The line $B+D=1$
thus acts like a mirror in the $(B,D)$-plane. The probability
$\Pi_1({\cal N}^*)$ along this line shows interesting corner-like
singularities where the most probable population makes a jump. In a
statistical mechanical context what we are doing is similar to
minimizing the ``free energy" $F = -\Pi_1({\cal N})$ with respect to
${\cal N}$, for every $B$, and obtaining the minimum-free-energy
(maximum-$\Pi_1$) curve which displays corners at points where two
``phases" (values of ${\cal N}^*$) coexist. Here, the relevant
${\cal N}$ are restricted to the seven population values generated
in the first generation (Fig.3a), so that the curve can have (at
most) six corners. Fig.6 shows the probability of the most probable
population along this line (case $B+D=1$). Although this curve has
been calculated only for the first generation, it is important to
keep in mind that it is already a good approximation to the
probability of the most probable carrying capacity $K^*$, defined in
the limit of a large number of generations. That curve, $\Pi(K^*)$
must also displays corner-like singularities, representing jumps in
$K^*$, at values of $B$ (or $D$) close to the singularities
exhibited in Fig.6. These jumps in $K^*$ occur when two main peaks
in the distribution (see Fig.3c for example) exchange maximum
height.

A global plot of the most probable population ${\cal N}^*$ after one
generation, in the domain $0 \leq B,D \leq 1$ is presented in Fig.7.
The two symmetries we discussed can be seen in this plot. The
singularities, in the form of terrace border lines, represent the
``coexistence" of two distinct values of $K^*$. These lines run
roughly parallel to lines of constant $B-D$, so that, to a crude
first approximation, the most probable population depends mainly on
the difference $B-D$. Recall that the average population, $\bar N$,
depends {\em only} on $B-D$, according to (II.1).

\begin{figure}[htbp]
\centerline{\epsfxsize=8cm \epsfbox{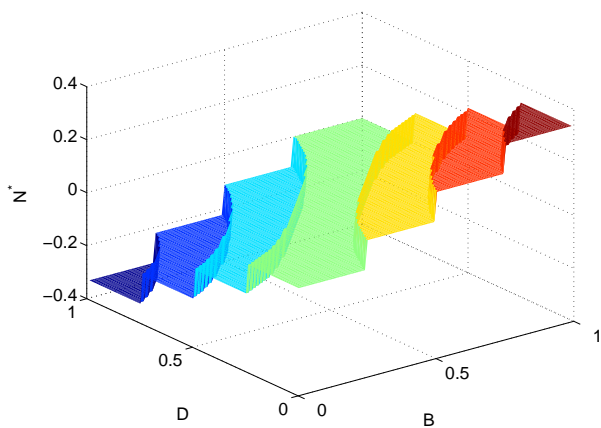}}
\caption{Most probable population after one generation as a
function of $B$ and $D$. The staircase structure reveals where two
values of $N^*$ exchange maximum probability. The model symmetries
are apparent and help interpret Fig.6 and Fig.8.}
\end{figure}

The probability $\Pi_1 ({\cal N}^*)$ as a function of $B$ and $D$ is
shown in Fig.8. It displays lines of corner-like singularities
precisely where the terrace borders occur in Fig.7. Note that a
sharp population distribution is only found near the four points
where $B$ and $D$ are close to 0 or 1.

\begin{figure}[htbp]
\centerline{\epsfxsize=8cm \epsfbox{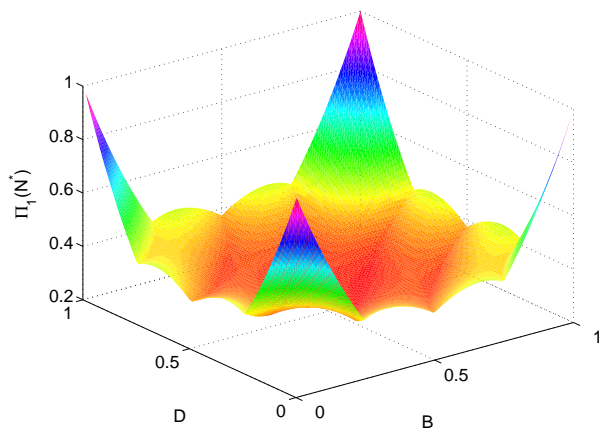}}
\caption{Probability of the most probable population after
one generation, as a function of $B$ and $D$. Two special lines
along this surface are shown in Fig.6. Notice that the border lines
at $B=0$ or $B=1$ display three corner singularities, as do the
lines for $D=0$ or $D=1$.}
\end{figure}

In order to quantify this further we have calculated the standard
deviation $\Sigma_1$ of the distribution after one generation, as a
function of $B$ and $D$. This quantity is defined through
\begin{equation}
(\Sigma_1)^2 = \frac{1}{2\lambda +1} \sum_{k=0}^{2\lambda} ({\cal
N}_k-\bar N(1))^2 \hat{\cal P}_1(k),
\end{equation}
with ${\cal N}_k \equiv 1/(\lambda-1) + (\lambda - k)/\lambda^2$
and, from (II.2), $\bar N(1) =1/(\lambda-1) + (B-D)/\lambda $.
Explicit expressions for the $\hat {\cal P}_1$ are given in the
Appendix. The standard deviation is shown in Fig.9. Like the average
population, the standard deviation is a smooth function of $B$ and
$D$, and illustrates clearly that a broad population distribution is
generic.

\begin{figure}[htbp]
\centerline{\epsfxsize=8cm \epsfbox{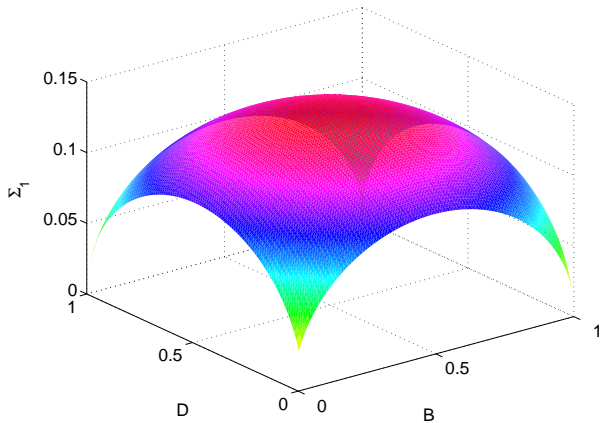}}
\caption{Standard deviation of the population distribution
after one generation, versus $B$ and $D$. Notice that only extreme
values of $B$ and $D$ (near the corners of the square) lead to a
sharp distribution and easily predictable populations of individual
samples.}
\end{figure}

\setcounter{equation}{0}
\renewcommand{\theequation}{\thesection.\arabic{equation}}
\section{Conclusions and a possible experimental test}
Keeping in mind the possible relevance of this hierarchical model to
towering pillar biofilms, as discussed in the Introduction, we
recapitulate here the main model ingredients. The model contains
three parameters, birth probability $B$, death probability $D$ and
rescaling factor $\lambda$. $B$ essentially reflects the fraction
(or quality) of nutrient-rich area on the surface, and $D$ the
density (or quality) of bactericidal agents. The rescaling factor
$\lambda$ sets the characteristic time scale (Section II) and
controls the fragmentation - in space - of the nutrient-rich patches
and the antimicrobial activity, mimicking diffusion and transport.
Further, the same $\lambda$ is used to control the saturation of the
population by decreasing uniformly the amount of offspring in each
generation and simultaneously increasing uniformly antimicrobial
resistance. This mimicks, for example, a gradual temperature drop
and biofilm development, respectively, as discussed in Section I.
Through the action of $\lambda$ the model becomes hierarchical and
this leads to a finite carrying capacity (``freezing" or
stagnation). Undoubtedly, the use of just one parameter $\lambda$
for inducing all these dynamical effects, in the interest of
simplicity and transparency of the model, cannot be more than a
first crude step towards a more realistic and refined approach.

The growth of the population thus acquires a hierarchical structure,
apparent in the fractal population ``landscapes" (Fig.1), and
expressed by a (nearly) self-similar carrying capacity distributions
(Figs.3-5), the mathematics of which has been discussed using
elementary notions of fractal geometry. The main conclusion from
studying these distributions is that for generic values of $B$ and
$D$ a broad range of carrying capacities can be observed, and the
outcome depends largely on the sample used. Ensemble averaging is
necessary for predicting statistically relevant properties of this
type of growth.

The question now arises whether this model is relevant to an
experiment in which bacteria capable of forming a biofilm on an
inhomogeneous nutrient field are exposed to antibiotic or protein
spray and subsequently put in a refrigerator or an oven (for about a
day, with $\Delta T \approx 1^\circ $C/hr). The applicability of the
model relies primarily on the determination of the model parameters
starting from experimental system parameters. In our discussion we
have taken $\lambda =3$, having in mind that the linear size of a
nutrient-rich patch or an antibiotic drop is roughly one third of
the inoculation line. This length ratio can be determined easily in
practice and the rescaling factor can be adjusted, or, of course,
two separate rescaling factors can be introduced if necessary. The
external cooling (or heating) rate, taking the bacteria away from
optimal growth conditions, is assumed to be adjusted so that the
processes come to a halt in about 4 to 5 generations. The quality of
the nutrient and the strength of the antibiotic, incorporated in $B$
and $D$, can be tuned experimentally by dilution, for example.

State-of-the-art experiments allow a simultaneous counting of many
samples (say, $10^4$) on a micro-array in order to obtain carrying
capacity distributions, much like is done in ensemble averaging. In
sum, a direct experimental check of this model is feasible, and
would be worthwhile, before introducing theoretical refinements
which undoubtedly are necessary to make the model more realistic,
but on the other hand compromise the insight that can be gained
using only very few key parameters.

{\bf Acknowledgments.}\\ We thank Ralf Blossey for suggesting the
possible applicability of the hierarchical population model to
biofilms. We are grateful to Reiner Kree and Joachim Krug for
constructive comments on the model. This research is supported by
the Flemish Programme FWO-G.0222.02 ``Physical and interdisciplinary
applications of novel fractal structures". K. S.-W. is supported by
the Foundation for Polish Science (FNP Scholarship).
\appendix
\setcounter{equation}{0}
\renewcommand{\theequation}{\thesection.\arabic{equation}}
\section{Probabilities of the population after one generation: case $\lambda =3$}
In this Appendix we give the analytic expressions used in the
calculations of the most probable population and its probability
after one generation (Section V). The rescaling factor is $\lambda
=3$. In terms of the auxiliary probabilities
\begin{equation}
P=B(1-D)\; \mbox{and} \; Q=D(1-B)
\end{equation}
with $0\leq B,D\leq 1$, and in terms of $S=P+Q$, we obtain, using
(III.4),
\begin{eqnarray}
\hat {\cal P}_1(0)& =& P^3 \nonumber \\ \hat {\cal P}_1(1)& =&
3P^2(1-S) \nonumber \\ \hat {\cal P}_1(2)& =& 3P(1-S)^2 + 3P^2Q
\nonumber \\ \hat {\cal P}_1(3)& =& (1-S)^3 +6P(1-S)Q \nonumber \\
\hat {\cal P}_1(4)& =& 3(1-S)^2Q +3PQ^2 \nonumber \\ \hat {\cal
P}_1(5)& =& 3(1-S)Q^2 \nonumber \\ \hat {\cal P}_1(6)& =& Q^3
\end{eqnarray}

\end{document}